\documentstyle[11pt]{article}
\begin{document}
\thispagestyle{empty}
\baselineskip 18pt
\rightline{UOSTP-99-007}
\rightline{SNUTP-99-042}
\rightline{KIAS-P99078}
\rightline{{\tt hep-th/9909035}}

\

\def\tr{{\rm tr}\,} \newcommand{\beq}{\begin{equation}}
\newcommand{\eeq}{\end{equation}} \newcommand{\beqn}{\begin{eqnarray}}
\newcommand{\eeqn}{\end{eqnarray}} \newcommand{\bde}{{\bf e}}
\newcommand{\balpha}{{\mbox{\boldmath $\alpha$}}}
\newcommand{\bsalpha}{{\mbox{\boldmath $\scriptstyle\alpha$}}}
\newcommand{\betabf}{{\mbox{\boldmath $\beta$}}}
\newcommand{\bgamma}{{\mbox{\boldmath $\gamma$}}}
\newcommand{\bbeta}{{\mbox{\boldmath $\scriptstyle\beta$}}}
\newcommand{\lambdabf}{{\mbox{\boldmath $\lambda$}}}
\newcommand{\bphi}{{\mbox{\boldmath $\phi$}}}
\newcommand{\bslambda}{{\mbox{\boldmath $\scriptstyle\lambda$}}}
\newcommand{\ggg}{{\boldmath \gamma}} \newcommand{\ddd}{{\boldmath
\delta}} \newcommand{\mmm}{{\boldmath \mu}}
\newcommand{\nnn}{{\boldmath \nu}}
\newcommand{\diag}{{\rm diag}}
\newcommand{\bra}[1]{\langle {#1}|}
\newcommand{\ket}[1]{|{#1}\rangle}
\newcommand{\sn}{{\rm sn}}
\newcommand{\cn}{{\rm cn}}
\newcommand{\dn}{{\rm dn}}

\

\vskip 0cm
\centerline{\Large\bf Comments on  the Moduli Dynamics of 1/4 BPS Dyons}

\vskip 0.2cm

\vskip 1.2cm
\centerline{\large\it
Dongsu Bak $^a$\footnote{Electronic Mail: dsbak@mach.uos.ac.kr}
and Kimyeong Lee $^{bc}$\footnote{Electronic Mail:
klee@kias.re.kr} }
\vskip 10mm
\centerline{ \it $^a$ Physics Department,
University of Seoul, Seoul 130-743, Korea}
\vskip 3mm
\centerline{ \it $^b$ Physics Department and Center for Theoretical
Physics}
\centerline{ \it Seoul National University, Seoul 151-742, Korea}
\vskip 2mm
\centerline{ \it $^c$ School of Physics, Korea Institute for Advanced
Study}
\centerline{ \it 207-43, Cheongryangryi-Dong, Dongdaemun-Gu, Seoul
130-012, Korea\footnote{address after 9-01-1999}}
\vskip 1.2cm
\begin{quote}
{\baselineskip 16pt We rederive the nonrelativistic Lagrangian for the
low energy dynamics of 1/4 BPS dyons  by considering
the time dependent fluctuations around classical 1/4 BPS
configurations.  The relevant fluctuations are the  zero modes of the
underlying 1/2 BPS monopoles.}
\end{quote}


\newpage

\baselineskip 13pt

Recently the 1/4 BPS dyonic configurations are constructed and their
nature has been exploited in the ${\cal N}=4$ supersymmetric
Yang-Mills theories~\cite{yi,hash,bak,kml}. Since the supersymmetric
Yang-Mills theories arise as a low energy description of parallel D3
branes in the type IIB string theory~\cite{witten}, the quantum 1/4
BPS states have the string interpretation as multi-pronged
string~\cite{bergman}.  In the classical field theory, the 1/4 BPS
configurations can be viewed as a collection of 1/2 BPS dyons
positioned with respect to each other so that a balance of the Coulomb
and Higgs forces is achieved.  The BPS equations satisfied by the
classical 1/4 BPS configurations consist of the 1/2 BPS monopole
equation and its gauge zero mode equation. The underlying 1/2 BPS
configurations are uniquely determined by the moduli coordinates,
which in turn determine the solution of the second BPS equation
uniquely~\cite{yi}.
  
The low energy dynamics of 1/4 BPS monopoles has been explored in
Ref.~\cite{leex} and it is shown that a specific potential is required
in addition to the kinetic terms over the moduli space. The basic
ideas of the construction were as follows.  In the limit where 1/4 BPS
configurations are almost 1/2 BPS, it should be possible to rediscover
the physics of 1/4 BPS configurations from the zero mode dynamics of
1/2 BPS configurations.  Since static forces exist between 1/2 BPS
solitons in the case of misaligned vacua~\cite{fraser}, the simplest
possibility is to add a potential term to the moduli space dynamics.
The potential is indeed uniquely determined from the given knowledge
of the electric charge and mass of the 1/4 BPS states. Here the result
by Tong was particularly useful~\cite{tong}. The low energy Lagrangian
has a BPS bound and its BPS configuration corresponds to the 1/4 BPS
field configuration~\cite{leex}. For a simple case, quantum 1/4 BPS
states of the corresponding supersymmetric Lagrangian have been found
in Ref.~\cite{leexx}.

However, the derivation of the low energy dynamics was in some sense
indirect. Even though the presence of the potential is obvious by
considering the interaction between point particle dyons, the exact
structure of the potential cannot be obtained from the 
particle point of
view.  In this note, we rederive the low energy Lagrangian for 1/4 BPS
configurations by the field theoretic method. The dynamical variables
are the zero modes, or the moduli of the underling 1/2 BPS
configurations.

We begin with the  ${\cal N}=4$  supersymmetric Yang-Mills 
theory. We choose the compact semisimple group $G$ of the rank $r$.
Among the six Higgs fields, only two Higgs fields $a, b$ play the role
in the BPS bound. The  bosonic part of the Lagrangian is given  by
\begin{equation}
L = \frac{1}{2} \int d^3x\; {\rm tr} \left\{ {\bf E}^2 -{\bf B}^2 +
(D_0a)^2 -({\bf  D} a)^2 + (D_0 b)^2 -({\bf D} b)^2 - \bigl(
-i[a,b]\bigr)^2 \right\}  ,
\label{lag}
\end{equation}
where $D_0 = (\partial_0 -i A_0) $, ${\bf D} = \nabla - i {\bf A}$, and
${\bf E} = \partial_0 \bf A - {\bf D} A_0$.
The four vector potential $(A_0, {\bf A}) = (A_0^a T^a, {\bf
A}^a T^a)$ and the group generators $T^a$ are traceless hermitian
matrices such that $\tr T^a T^b=\delta^{ab}$. 

As shown in Ref.~\cite{yi},  there is a BPS bound on the energy 
functional, which is saturated when  configurations satisfy
\begin{eqnarray}
&& {\bf B}={\bf D}b\,,
\label{bogo1} \\
&&{\bf E}={\bf D} a \,,\ \
\label{bogo2} \\
&&
D_0 b-{i}[a, b]=0\,, \label{bogo3} \\
&& 
D_0 a=0\,,\ \  \label{bogo4}
\end{eqnarray}  
together with the Gauss law,
\begin{equation}
\label{gauss}
{\bf D}\cdot {\bf E} -i[b,D_0 b]-i[a,D_0 a]=0\,. 
\end{equation}
Equation (\ref{bogo1}) is the old BPS equation for 1/2 BPS
monopoles and is called the primary BPS equation. Equations
(\ref{bogo2}), (\ref{bogo3}), (\ref{bogo4}), and (\ref{gauss}) can 
be put together into a single equation, 
\begin{equation}
{\bf D}^2 a -[b,[ b, a]]=0 , 
\label{secondary}
\end{equation} 
which is called  the secondary BPS equation. This equation is 
the global gauge zero mode equation for the first
BPS equation.  We can choose the gauge where $A_0 = -a$, in which case
the configuration itself becomes static in time.

In the asymptotic region, two Higgs fields take the form
\begin{eqnarray}
&& b \simeq {\bf b}\cdot {\bf H} - \frac{{\bf g}\cdot {\bf H}}{4\pi r}, \\
&& a \simeq {\bf a}\cdot {\bf H} - \frac{{\bf q}\cdot {\bf H}}{4\pi
r},
\label{aasympt}
\end{eqnarray}
where ${\bf H}$ is the Cartan subalgebra.  We are interested in the
case where the expectation value ${\bf b}$ breaks the gauge group $G$
maximally to abelian subgroups $U(1)^r$.  Then, there exists a unique set
of simple roots $\betabf_1,\betabf_2,...,\betabf_r$ such that
$\betabf_\alpha \cdot {\bf b} > 0$~\cite{ejw}. The magnetic and
electric charges are given by
\begin{eqnarray}
&& {\bf g} = 4\pi \sum_{\alpha=1}^{r} n_\alpha \betabf_\alpha ,\\
&& {\bf q} = \sum_{\alpha=1}^{r} q_\alpha \betabf_\alpha ,
\end{eqnarray}
where integer $n_\alpha \ge 0$.  Any solution to these BPS equations
possesses a mass that saturates the BPS bound
\begin{equation}
\label{mass}
M=Z= {\bf b} \cdot {\bf g} + {\bf a} \cdot {\bf q} ,
\end{equation} 
where $Z$ is the larger  one out of two central 
charges in the ${\cal N}=4$ supersymmetric theory.

The solutions of the primary BPS equation describe the collection of
1/2 BPS monopoles. For each simple root, there exists a fundamental
monopole of four zero modes. The integer $n_\alpha$ denotes the number
of the $\betabf_\alpha$ fundamental monopoles.  We consider the case
where all $n_\alpha$ are positive so that the monopoles do not separate
into mutually noninteracting subgroups.  The moduli space of the 1/2
BPS configuration has the dimension of the number of zero modes,
$4\sum_\alpha n_\alpha$. With the moduli space coordinates $z^M$, the
zero modes are a linear combination of moduli coordinate dependence
and a local gauge transformation. With a simple pseudo four
dimensional vector $A_\mu({\bf x}, z^M) = ({\bf A}, b)$ with
$\mu=1,2,3,4$, the zero modes will be
\begin{equation}
\delta_M A_\mu = \frac{ \partial A_\mu }{\partial z^M }+ D_\mu
\epsilon_M,
\end{equation}
where $D_\mu \epsilon_M = \partial_\mu \epsilon_M - i [A_\mu, \epsilon_M]$ with
understanding $\partial_4=0$. The zero mode equations for the primary
BPS equation are
\begin{eqnarray}
&& \nabla \times \delta_M {\bf A} = \nabla \delta_M b - i [ \delta_M {\bf
A}, b] ,\\
&& D_\mu \delta_M A_\mu = 0,
\label{background}
\end{eqnarray}
where the second equation is the background field gauge fixing
condition. {}From the field theory, there is well defined metric on the
moduli space~\cite{manton,atiyah,blum}, 
\begin{equation}
g_{MN}(z) = \int d^3 x \; {\rm tr} \delta_M A_\mu \delta_N A_\mu.
\end{equation}
The  low energy dynamics of 1/2 BPS configurations
is given by the nonrelativistic Lagrangian
\begin{equation}
L_{1/2} = \frac{1}{2} g_{MN}(z)\dot{z}^M \dot{z}^N.
\label{lag12}
\end{equation}

As there are $r$ unbroken global $U(1)$ symmetries, the corresponding
electric charges should be conserved. In another word, $L_{1/2}$
should have $r$ cyclic coordinates corresponding to these gauge
transformations. For each $\betabf_\alpha$ $U(1)$ symmetry the
corresponding cyclic coordinate is denoted by $\psi^\alpha$ with
$\alpha=1,...,r$.  Expanding the asymptotic value ${\bf a} =
\sum_\alpha a^\alpha \lambdabf_\alpha$, where $\lambdabf_\alpha$'s are
the fundamental weights such that $\lambdabf_\alpha \cdot \betabf_\beta
= \delta_{\alpha\beta}$, we notice $D_\mu a$ is the gauge zero mode
\begin{equation}
D_\mu a = a^\alpha K_\alpha^M \delta_M A_\mu,
\end{equation}
where 
\begin{equation}
K^M_\alpha \frac{\partial}{\partial z_M} = \frac{\partial}{\partial
\psi_\alpha} 
\end{equation}
is the Killing vector for the $\betabf_\alpha$ $U(1)$ symmetry. 
If we divide the moduli coordinates ${z^M}$ to ${\psi^\alpha}$ and the
rest $y^i$, the Lagrangian (\ref{lag12}) can be rewritten as
\begin{equation}
L_{1/2} = \frac{1}{2} h_{ij}(y) \dot{y}^i \dot{y}^j 
+ \frac{1}{2} L_{\alpha\beta}(y) (\dot{\psi}^\alpha + w^\alpha_i(y)
\dot{y}^i) (\dot{\psi}^\beta + w^\beta_j(y)
\dot{y}^j).
\end{equation}
Here  $h_{ij} = g_{ij}$, $L_{\alpha\beta} = g_{MN} K_\alpha^M K_\beta^N$,
and $w^\alpha_i = L^{\alpha\beta} g_{\beta i}$.
Notice that all metric components are independent of $\psi^\alpha$.

Let us now explore the low energy dynamics of 1/4 BPS configurations.
The idea is to calculate the field theoretic Lagrangian for a
suitable initial condition in the field theory. It needs to specify
the fields and their time derivatives or their momenta. Clearly we
require the initial condition to be given by a 1/4 BPS configuration when
there is no real time evolution. As the momentum variables ${\bf E}$
and $D_0 b$ are nonzero for 1/4 BPS configurations, nontrivial time
evolution will ensue only if we add additional field momenta or time
derivatives to the 1/4 BPS configuration.

The moduli space dynamics of 1/2 BPS configurations is correct when
the kinetic energy is much smaller than the rest mass. This means the 
order of the velocities, $v \sim \dot{z}^M$ is much smaller than 1. 
For 1/4 BPS configurations, there is a natural scale $\eta \sim |{\bf
a}|/|{\bf b}|$. We will see that the limit  $\eta << 1$ is the suitable
region for the low energy dynamics.

Thus, let us put the initial condition to be ${\bf A}({\bf x},y^i),
b({\bf x},y^i), a({\bf x}, y^i)$ and the momentum variables, 
${\bf D}a + \dot{z}^M \delta_M {\bf A}$, $i[a,b] + \dot{z}^M \delta_M b$, and
$\dot{z}^M \delta_M a$. Here we have replaced the zeroth order momentum
variables with the field variables by using the 1/4 BPS equations. We
also choose the gauge $A_0=-a$.  $\delta_M a$ cannot be defined by
the zero mode equation of the secondary BPS equation. Otherwise the
asymptotic form (\ref{aasympt}) implies nonzero contribution from
$\partial_0 {{\bf q}(z)}$ to $\partial_0 a$. The 1/4 BPS condition
involves the field momenta and we cannot include the additional field
momenta at a given point of the moduli space, 
maintaining the 1/4 BPS equations. Rather we choose
$\dot{z}^M \delta_M a$ to be an unspecified quantity of order $\eta v$,
whose exact nature, as we will see soon, is irrelevant for the low
energy dynamics.  As $\delta_M {\bf A}$ and $\delta_M b$ satisfy the
background gauge, the Gauss law is satisfied for the initial condition
to order $v$. There is a correction of order $\eta^2 v$ due to the $a$
field, but it is negligible to the order we are working on.

Let us now calculate the Lagrangian (\ref{lag}) for this initial
condition. It becomes
\begin{equation}
L = -{\bf b}\cdot {\bf g} +\frac{1}{2} 
 \int d^3 x \, {\rm tr} \left\{ (\dot{z}^M
\delta_M  A_\mu)^2 \right\} + \int d^3 x \; \tr\left\{   \dot{z}^M
 \delta_M {\bf A} \cdot   {\bf D} a +  \dot{z}^M \delta_M b\;\, i [a,b] 
 \right\}  ,
\end{equation}
where the first order in velocity terms appear as there were nonzero
field momenta for the 1/4 BPS configurations. Here again we used 1/4
BPS equations to replace the momenta with the fields.  This Lagrangian
is of order $v^2$ or $\eta v$. We have neglected the terms of order
$v^2 \eta^2$, which comes from the kinetic energy of $a$ field.  The
terms linear in $\dot{z}^M$ can be rewritten as a boundary
contribution by using the background gauge condition,
\begin{equation}
{\dot{z}}^M \int d^3x \; \tr \bigl\{ \delta_M {\bf A}\cdot {\bf D} 
a + \delta_M
 b \; \, i [a,b]
\bigr\} = {\dot{z}}^M \int d^3 x \; \nabla \cdot \tr (  a \delta_M
{\bf A}) .
\end{equation}
Noticing that ${\bf D} a$ and $i[a,b]$ are a  global gauge zero
mode, $ a^\alpha K_\alpha^M \delta_M A_\mu$, we can rewrite the
nonrelativistic Lagrangian as
\begin{equation}
L_1= \frac{1}{2}  g_{MN}\dot{z}^M \dot{z}^N + g_{MN} \dot{z}^M a^\alpha
K_\alpha^N
\label{lag1}
\end{equation}
where $-{\bf b}\cdot {\bf g}$ is omitted.

Let us introduce new moduli coordinates $\{ \zeta^M\} = {y^i,
\chi^\alpha}$ such that  
\begin{equation}
\chi^\alpha = \psi^\alpha +a^\alpha t .
\end{equation}
%
Since this transformation
shifts only cyclic coordinates, the above Lagrangian becomes
\begin{equation}
L_{1/4}  = \frac{1}{2} g_{MN}(\zeta) \dot{\zeta}^M \dot{\zeta}^N -
\frac{1}{2} g_{MN}(\zeta) a^\alpha K^M_\alpha a^\beta K_\beta^N.
\label{lag14}
\end{equation}
The kinetic term of this Lagrangian is the low energy Lagrangian
(\ref{lag12}) for 1/2 BPS configurations and there is an additional
potential. In terms of the $y^i, \chi^\alpha$ variables,
\begin{equation}
L_{1/4} =  \frac{1}{2} h_{ij}(y) \dot{y}^i \dot{y}^j 
+ \frac{1}{2} L_{\alpha\beta}(y) (\dot{\chi}^\alpha + w^\alpha_i(y)
\dot{y}^i) (\dot{\chi}^\beta + w^\beta_j(y)
\dot{y}^j)  - \frac{1}{2} L_{\alpha\beta}(y)  a^\alpha a^\beta .
\end{equation}
This is exactly the low energy Lagrangian obtained in Ref.~\cite{leex}. 

There is a  couple of more points to be discussed. The exact 1/4 BPS
configuration is static in $y^i$ and $\psi^\alpha$ coordinates, so
that $\chi^\alpha = a^\alpha t + {\rm constant\; term}$. The velocity
of $\chi^\alpha$ coordinates is of order $\eta$, which is all right as $v
$ and $\eta$ can be of the same order. When we define the Hamiltonian,
the $z^M$ coordinates are not appropriate. Again from the field
theory, the energy function we have has the contribution from the
momentum variables. In terms of $z^M$ variables, the field theoretic
energy functional for our initial condition becomes
\begin{equation}
E = {\bf b}\cdot {\bf q} + {\bf a}\cdot {\bf q} + L_1 .
\end{equation}
In terms of the $\{\zeta^M \}= \{y^i,
\chi^\alpha\}$ variables, this  energy becomes
\begin{equation}
E = {\bf b}\cdot{\bf q} + E_{1/4} ,
\end{equation}
where $E_{1/4}$ is the energy corresponding to the Lagrangian
(\ref{lag14}), 
\begin{equation}
E_{1/4}=  \frac{1}{2} g_{MN}(\zeta) \dot{\zeta}^M \dot{\zeta}^N +
\frac{1}{2} g_{MN}(\zeta) a^\alpha K^M_\alpha a^\beta K_\beta^N.
\end{equation}
Here we have used the Tong formula
\begin{equation}
{\bf a}\cdot {\bf q} = g_{MN}(\zeta)  a^\alpha K^M_\alpha a^\beta
K_\beta^N.
\end{equation}
The energy $E_{1/4}$ has a BPS bound, which  is saturated when
$\dot{z}^M=0$ or $\dot{\zeta}^M = a^\alpha K_\alpha^M$, and has the  value
of the electric mass ${\bf a}\cdot{\bf q}$. This nonrelativistic BPS
configuration describes the field theoretic 1/4 BPS configurations.
Thus, a consistent picture of the moduli space has been emerged.

As shown in the Ref.~\cite{leex}, the above Lagrangian can be
generalized to supersymmetric case so that it describes 1/4 BPS dyons
in the $N=4$ supersymmetric Yang-Mills theory. There exists naturally
 quantum BPS bound on this supersymmetric Lagrangian. In
Ref.~\cite{leexx}, the quantum 1/4 BPS states are found by solving the
quantum BPS conditions on the wave functions for
the case of the $SU(3)$ group.

\centerline{\bf Acknowledgments}

Part of this work is accomplished during our stay  
for Particles Fields and Strings '99 Conference of Pacific Institute
for the Mathematical Sciences at Vancouver, for which we
thank the hospitality of the center. 
We acknowledge a useful discussion with Piljin Yi.  D.B. is supported
in part by Ministry of Education Grant 98-015-D00061.  K.L.  is
supported in part by the SRC program of the SNU-CTP and the Basic
Science and Research Program under BRSI-98-2418.  D.B. and K.L. are
also supported in part by KOSEF 1998 Interdisciplinary Research Grant
98-07-02-07-01-5.

\end{document}